\documentclass[twocolumn,showpacs,preprintnumbers,amsmath,amssymb]{revtex4}

\usepackage{graphicx}

\begin{document}

\title{How to produce new superheavy nuclei?}

\author{K. Siwek-Wilczy\'nska}
\affiliation {Faculty of Physics, Warsaw University, Pasteura 5, 02-093 Warsaw, Poland}

\author{T. Cap}
\affiliation{National Centre for Nuclear Research, Ho\.za 69,
PL-00-681 Warsaw, Poland}
\author{M. Kowal}
\affiliation{National Centre for Nuclear Research, Ho\.za 69,
PL-00-681 Warsaw, Poland}
\date{\today}
\begin{abstract}
Existing experimental facilities limit the possibilities for discovery of new
nuclides to those synthesized with cross sections above 100 fb, but the perspectives for
future high current accelerators could lower this limit by two orders of magnitude.
Therefore, in the present work excitation functions for
fusion-$xn$ evaporation reaction channels induced not only by $^{48}Ca$
but also by heavier projectiles (usually leading to smaller cross sections) on actinide targets
were calculated in the framework of the fusion-by-diffusion (FBD) model.
For the first time, in this approach, channels in which a proton ($pxn$) or alpha particle ($\alpha$$xn$) is evaporated have been included in the first step of the deexcitation cascade.
To calculate the synthesis cross sections entry data such as fission barriers, ground-state masses, deformations and shell effects of the
superheavy nuclei calculated in a consistent way within the Warsaw macroscopic-microscopic model were used.
The only adjustable parameter of the FBD model is the injection point distance $s_{inj}$ and the value determined in our previous analysis of experimental cross sections for the synthesis of
superheavy nuclei of Z=114-118 has been used. Excitation functions for the synthesis of
selected (cross section above a few fb) new superheavies in the range of atomic
numbers 112-120 are presented. Observation of 21 new heaviest isotopes is predicted.
A realistic discussion of the FBD model uncertainties is presented for the first time.

\end{abstract}

 \maketitle

\section {Introduction}

The Fusion by Diffusion (FBD) model was proposed by W. J. \'Swi\c{a}tecki et al.
\cite{Acta,FBD-05} as a simple tool to calculate cross sections and optimum
bombarding energies for a class of reactions leading to the synthesis of superheavy nuclei. As in other theoretical models, in the FBD model the partial
evaporation-residue cross section for the synthesis of superheavy nuclei,
$\sigma_{ER}(l)$, is factorized as the product of the partial capture cross
section $\sigma_{cap}(l) =\pi\lambdabar^2(2l + 1)T (l)$, the fusion probability
$P_{fus}(l)$, and the survival probability $P_{surv}(l)$.
\begin{equation}
\label{factorize}
 \sigma_{ER} = \pi \lambdabar^2 \sum_{l = 0}^{\infty}(2l+1)
 T(l)\cdot P_{fus}(l)\cdot P_{surv}(l).
\end{equation}
Here, $\lambdabar$ is the wavelength, $\lambdabar^2=\hbar^2/2\mu E_{c.m.}$,
and $\mu$ is the reduced mass of the colliding system.

The key assumption which allows us to investigate the reaction mechanism in such a way is Bohr's hypothesis,
which states that the whole reaction process is a Markow type stochastic process which means that there are no memory effects.
This implies that the exit channel is completely independent of  the intermediate stage leading to the compound nucleus
as well as of the entrance channel. This hypothesis is justified by the different time scale of the particular reaction stages.

The capture transmission
coefficients $T(l)$ are calculated in a simple sharp cut off approximation,
where the upper limit $l_{max}$ of full transmission, $T(l)=1$, is determined
by the capture cross sections, known from the systematics described in Ref.
\cite{KSW04}.

The second factor, the fusion probability $P_{fus}(l)$, is the probability that
after reaching the capture configuration, the colliding system will eventually
overcome the saddle point and fuse, avoiding reseparation.  For very heavy and
less asymmetric systems, $P_{fus}(l)$ is much smaller than 1 and thus is
mainly responsible for the dramatically small cross sections for the production of
superheavy nuclei. The fusion hindrance in these reactions is caused by the
fact that for heaviest compound nuclei the saddle configuration is more
compact than the configuration of the two initial nuclei at sticking. It is
assumed in the FBD model that after sticking, a neck between the two nuclei
grows rapidly at an approximately fixed mass asymmetry and constant length of
the system \cite{Acta,FBD-05} bringing the system to the "injection point"
somewhere along the bottom of the asymmetric fission valley. To overcome the
saddle point and fuse, the system must climb uphill from the injection point to
the saddle in a process of thermal fluctuations in the shape degrees of freedom.
It was shown in Ref. \cite {Acta}  by solving the Smoluchowski diffusion
equation that the probability that a system injected on the outside of the saddle
point at an energy $H$ below the saddle point will achieve fusion is:
\begin {equation}
P_{fus} = \frac{1}{2}(1-{\rm erf}\sqrt{H/T}\,)
\end {equation}
where $T$ is the temperature of the fusing system.

The last factor in Eq. (1), $P_{surv}(l)$, is the probability for the compound
nucleus to decay to the ground state of the residual nucleus via evaporation of
light particles (neutrons, protons or alphas) and finally gamma deexitation
and thus avoid fission (survive). To calculate the survival probability
$P_{surv}$, the standard statistical model was used by applying the Weisskopf
formula for the particle emission width and the standard expression of the
transition-state theory for the fission width. The level density parameters for the
particle evaporation  channels were calculated as proposed by Reisdorf
\cite{Reisdorf} with shell effects accounted for by the Ignatyuk formula
\cite{Ignatyuk}. All details can be found in Ref. \cite{FBD-11}.

As follows from the above description, cross section calculations require
knowledge of the individual characteristics of the synthesized compound nuclei and their
decay products, all along the decay chain. The fission barriers, ground-state
masses, deformations and shell corrections of the superheavy nuclei predicted
using the Warsaw macroscopic-microscopic model were used \cite{MK,MK1}.

The only adjustable parameter of the FBD model is the injection point distance,
$s_{inj}$, defined as the excess of length of the deformed system at the
injection point configuration over the sum of the target and projectile
diameters. Its value was calculated from the systematics  determined in
our previous analysis of experimental cross sections for the synthesis of
superheavy nuclei of Z=114-118 \cite{KSW12}.

\begin{figure}[h!]
\includegraphics[width=5cm]{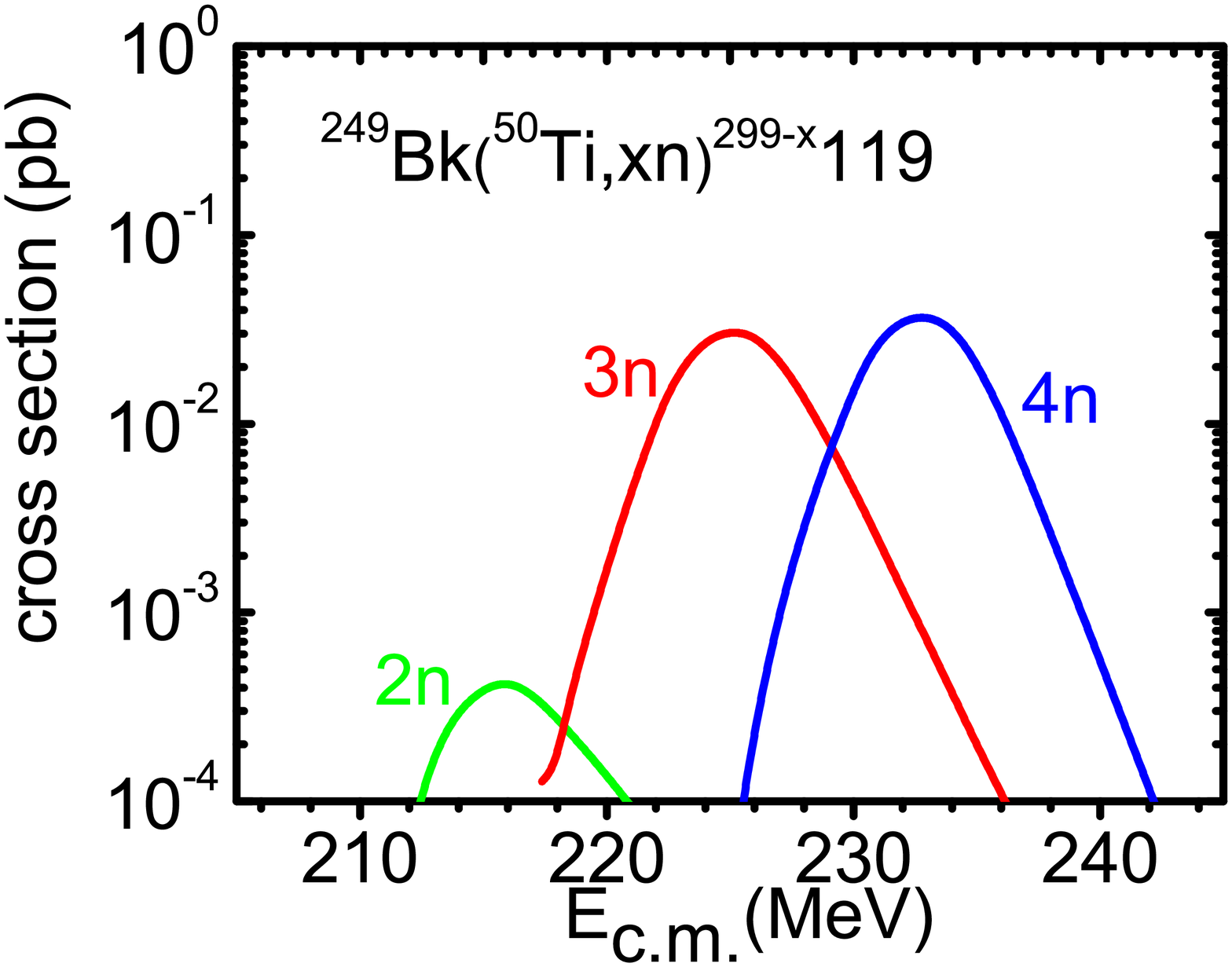}
\includegraphics[width=5cm]{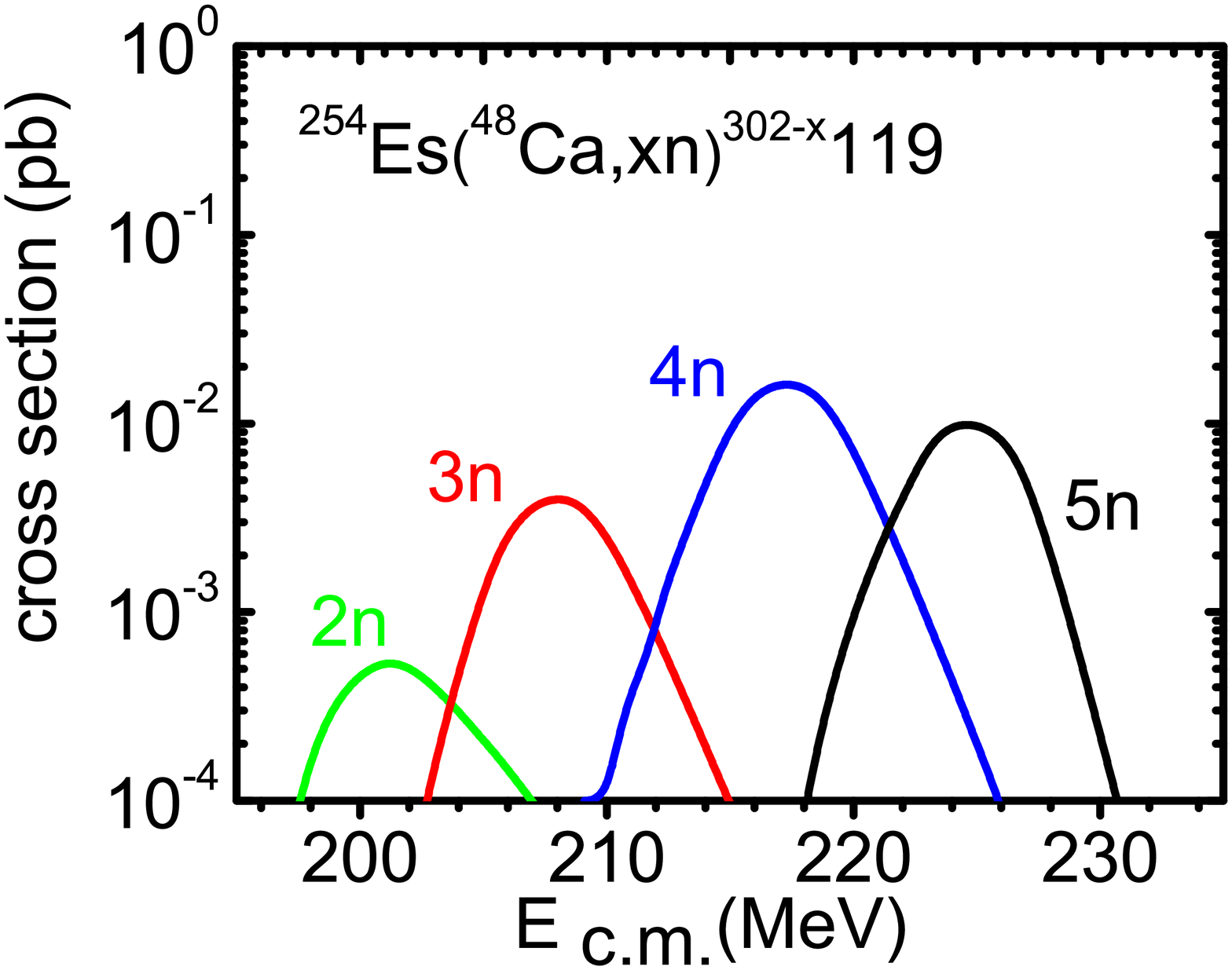}
\includegraphics[width=5cm]{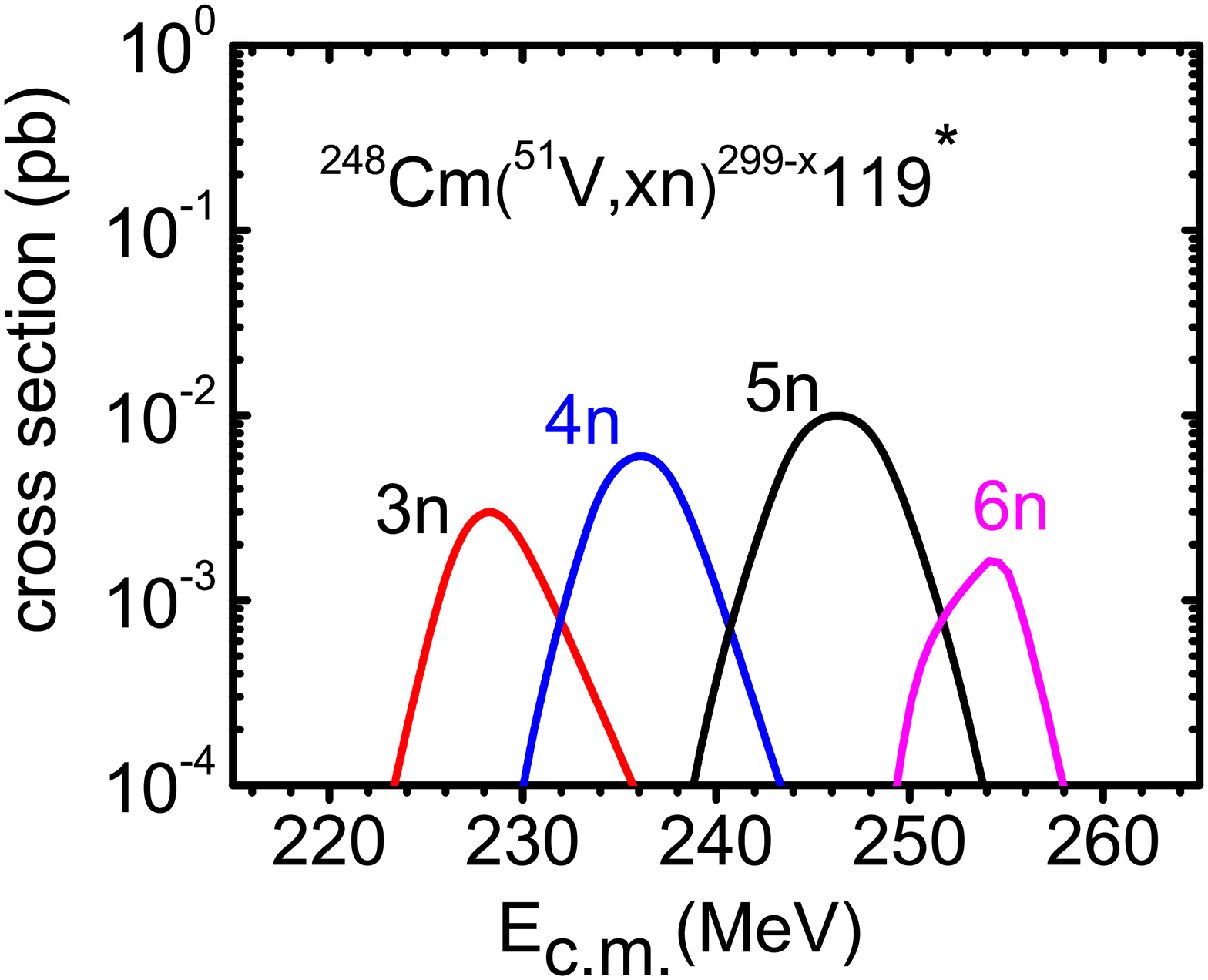}
\includegraphics[width=5cm]{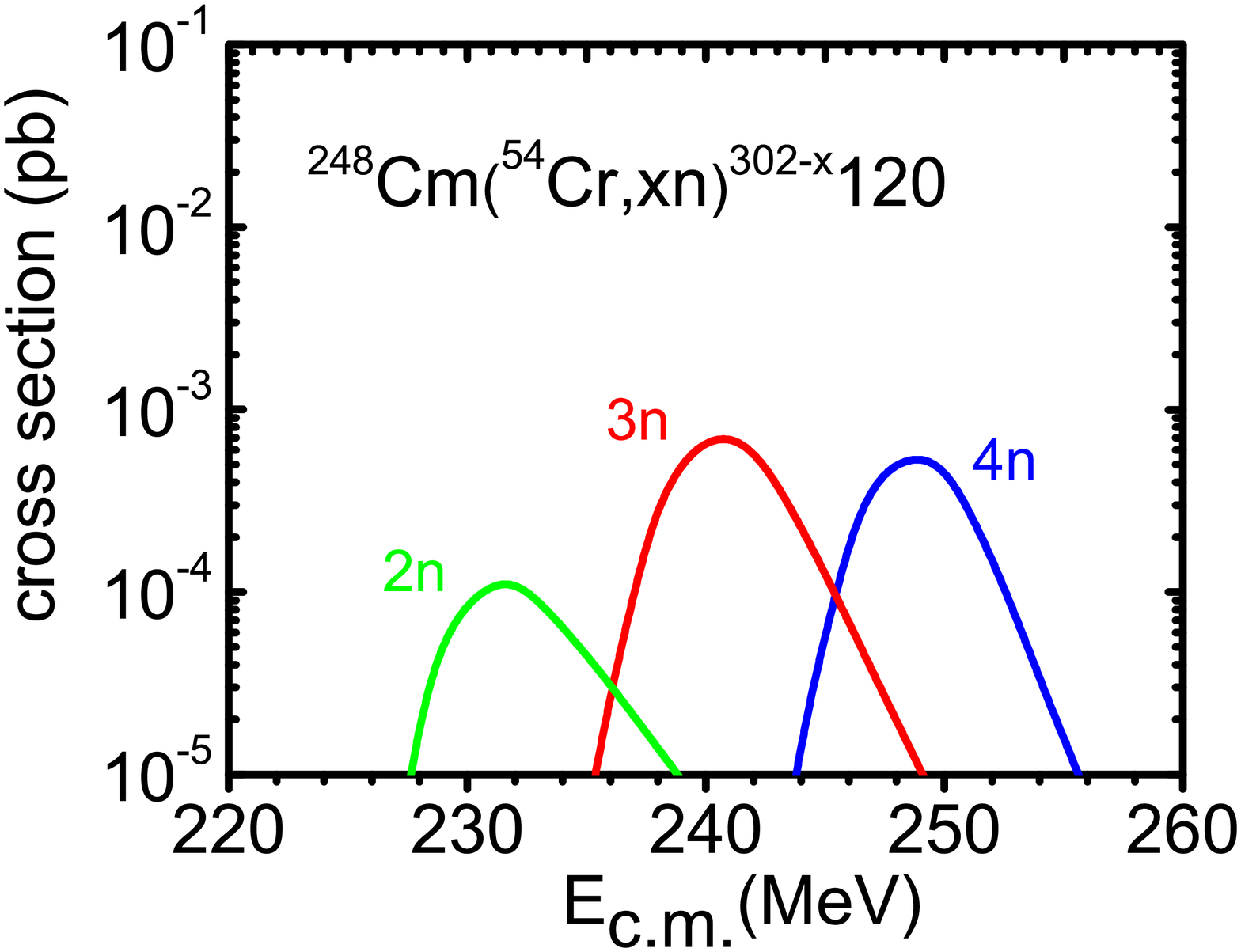}
\caption{(Color online) Cross sections for  the synthesis of superheavy nuclei of atomic
number Z = 119 and 120 predicted in the fusion-by-diffusion (FBD) model with
the fission barriers and ground-state masses of Kowal et al. \cite {MK, MK1} and
the systematics of the injection-point distance  (see text).} \label{Fig.1}
\end {figure}

\begin{figure}[t!]
\includegraphics[width=5cm]{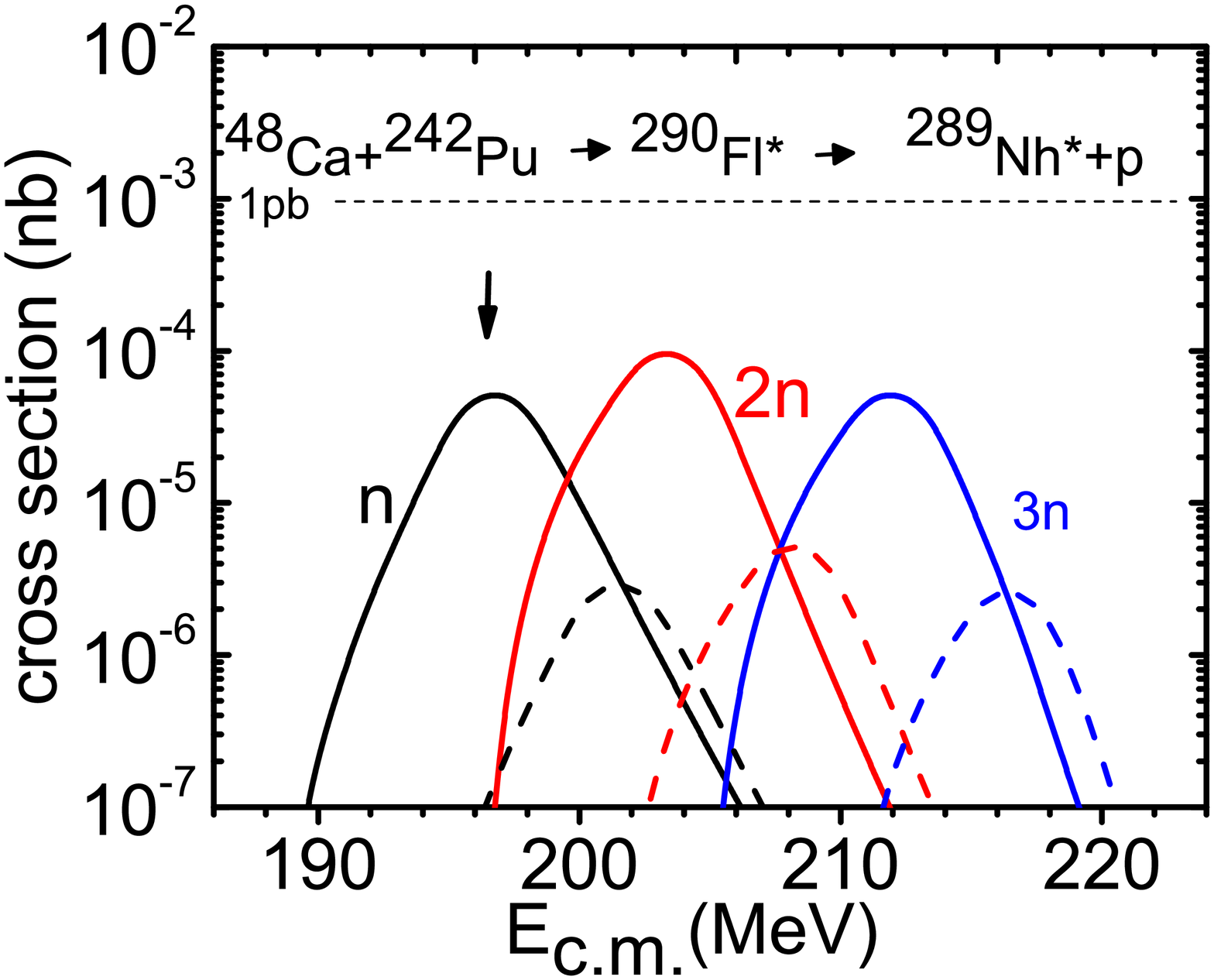}
\includegraphics[width=5cm]{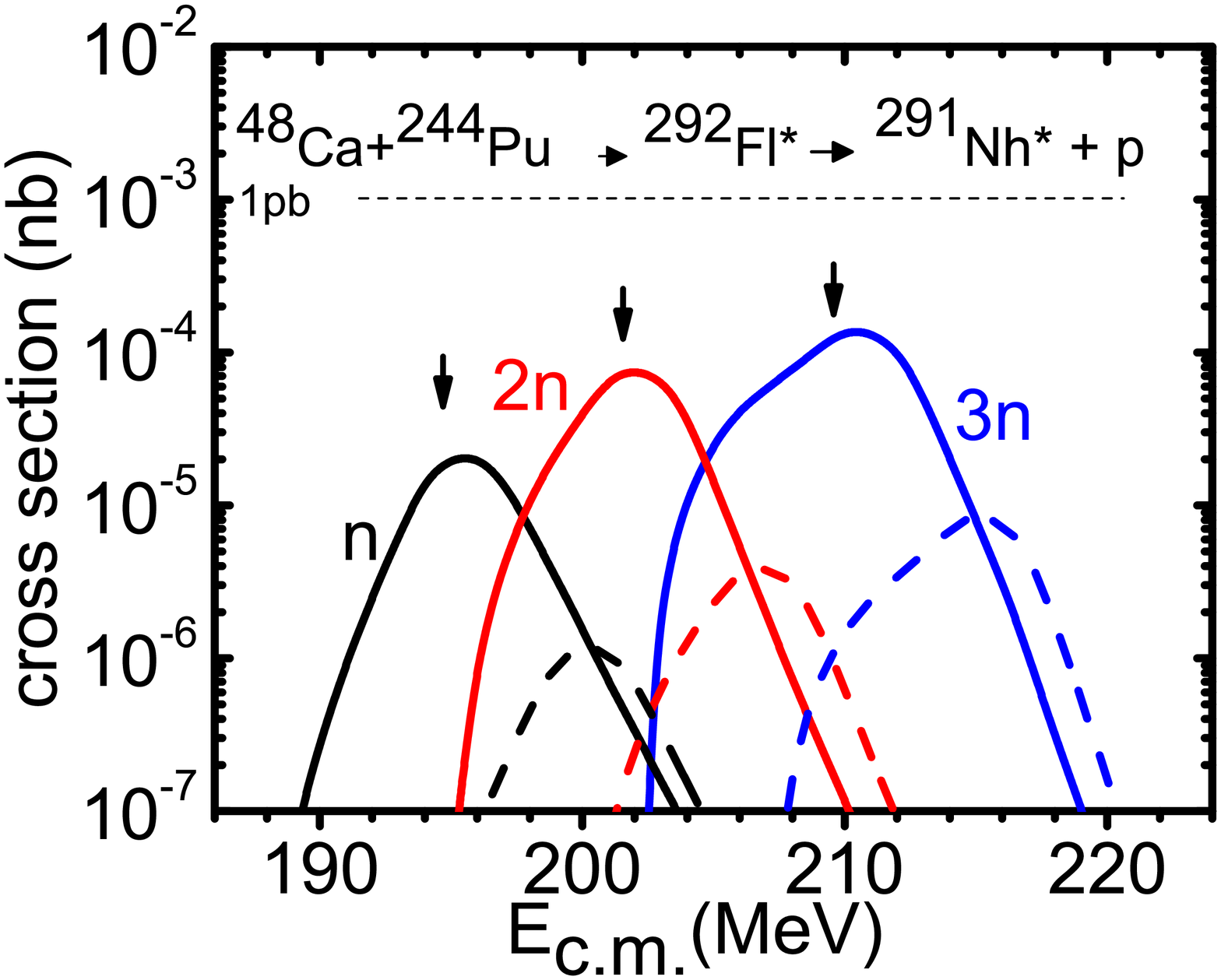}
\includegraphics[width=5cm]{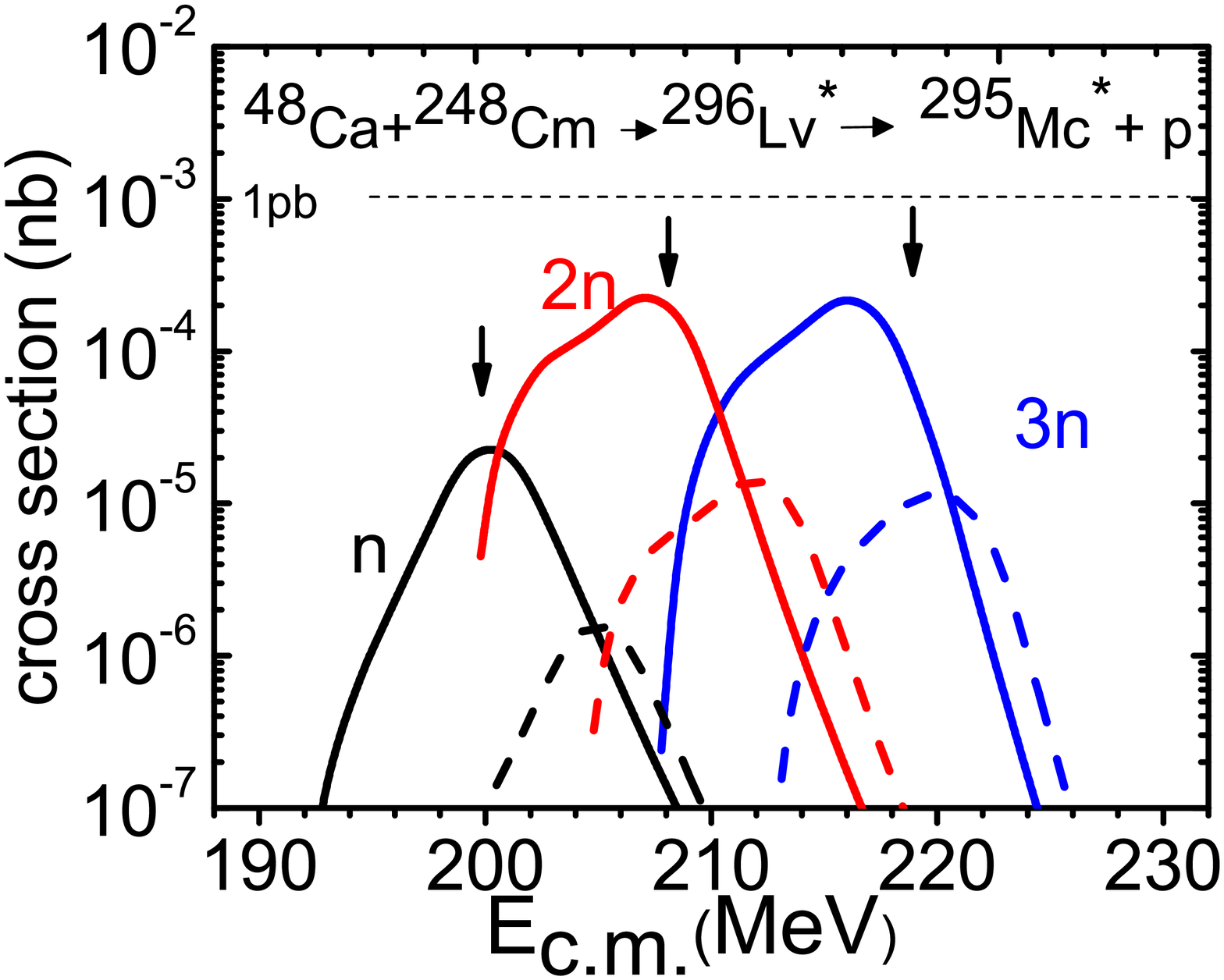}
\includegraphics[width=5cm]{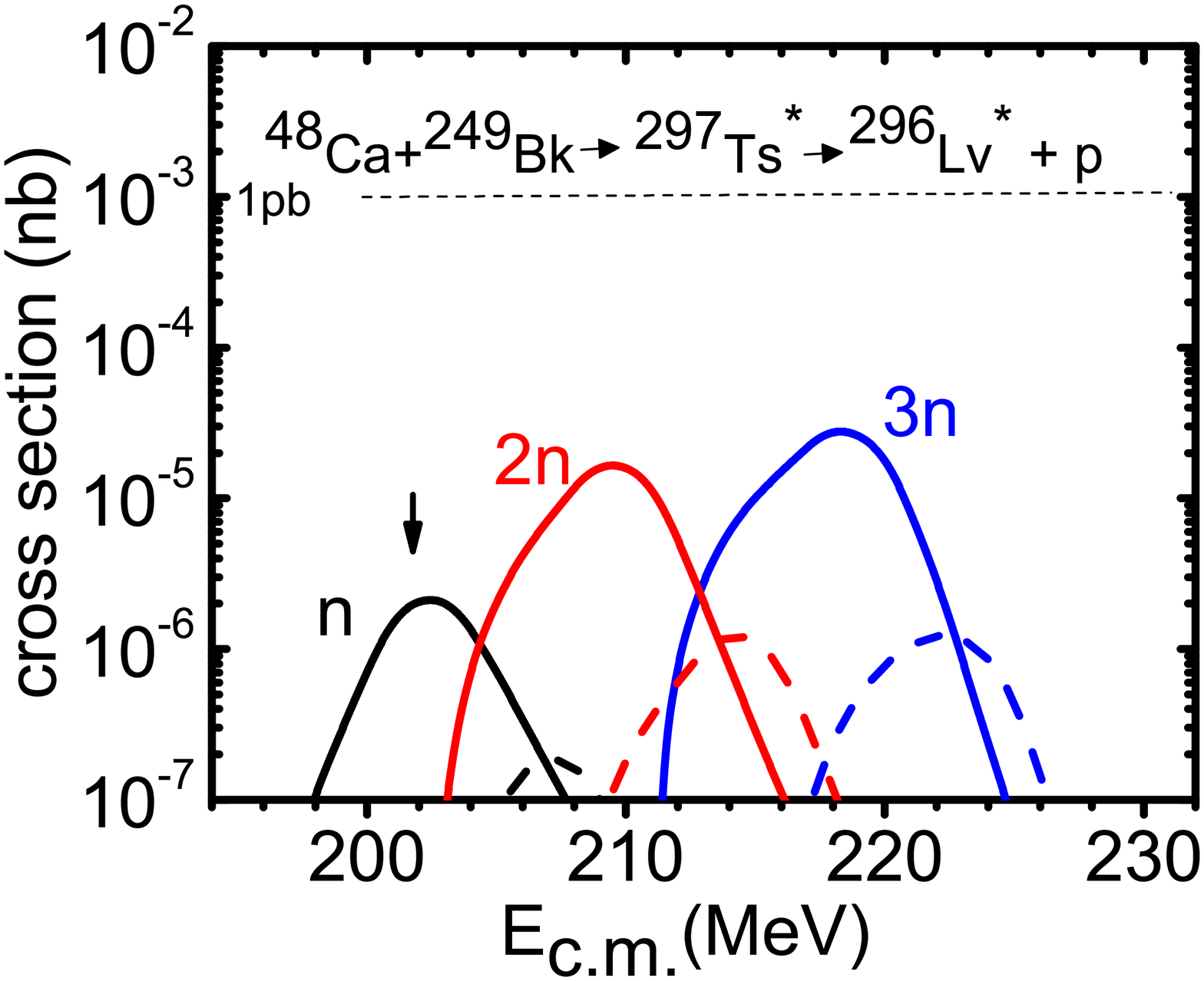}
\includegraphics[width=5cm]{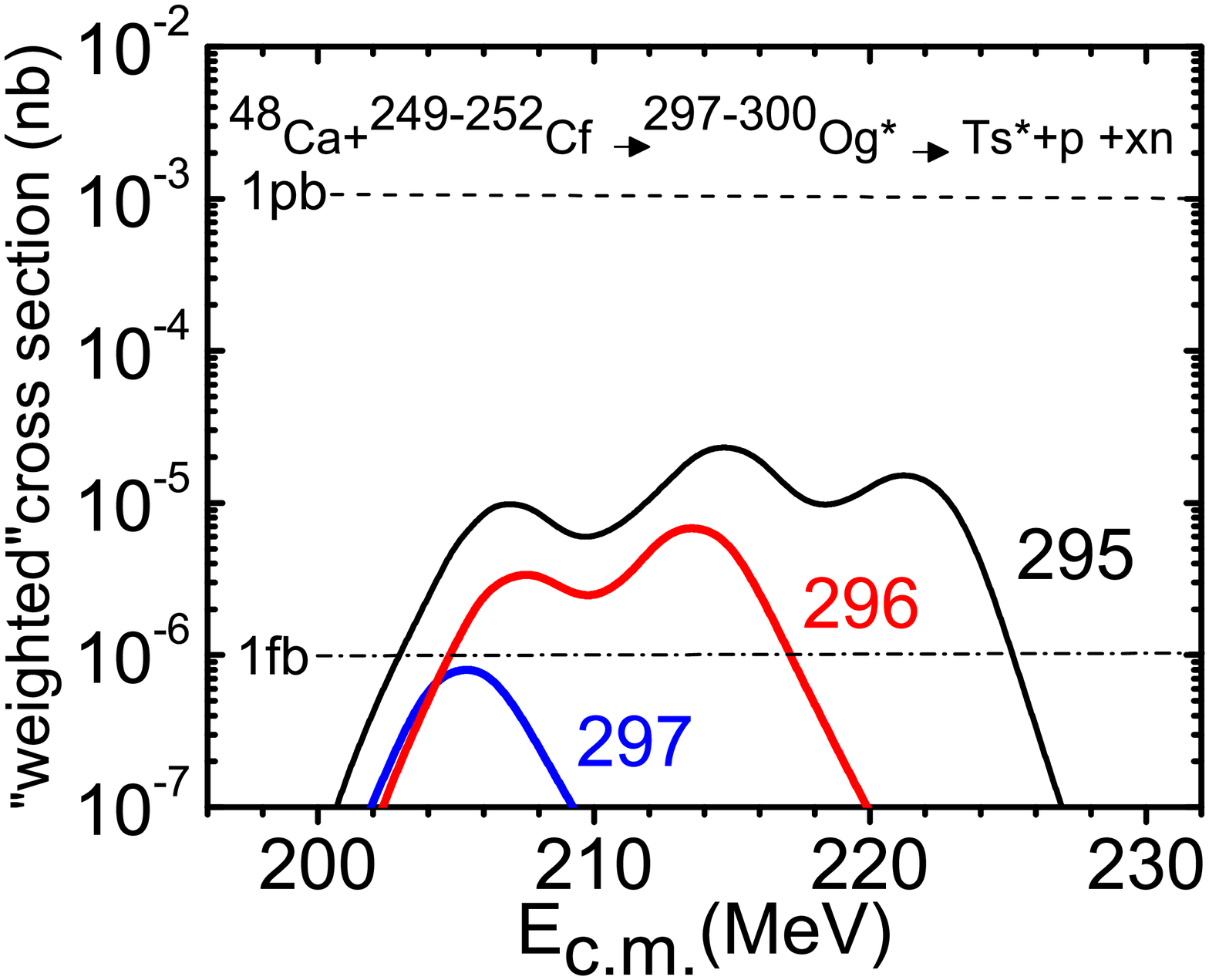}
\caption{(Color online) Cross sections for  the synthesis of superheavy nuclei in $pxn$ fusion evaporation processes, predicted by the fusion-by-diffusion (FBD) model with
the fission barriers and ground-state masses of Kowal et al. \cite{MK,MK1} and the
systematics of the injection-point distance (see text). The weigthed cross section takes into account the isotopes content of the mix californium target.} \label{Fig.2}
\end{figure}

\section {RESULTS}
\subsection{New elements} 
To synthesize new elements: Z=119 and 120 in $^{48}Ca$ induced fusion-$xn$ evaporation
reactions targets of $Es$ or $Fm$ are required respectively. Since they are
not currently available, reactions with heavier projectiles are
also considered here.
In Fig. 1 excitation functions for
 $^{50}Ti+^{249}Bk$, $^{48}Ca+^{254}Es$,
  $^{51}V+^{248}Cm$ and $^{54}Cr+^{248}Cm$ (predicted using the FBD model) are presented. Calculations for the above mentioned systems were also performed using other
models, see eg. \cite {Wang,Liu,Zhu,Dev,Umar,San,Fan} and citations there in. These cross sections are at least one order of magnitude smaller than cross
sections for the production of lighter superheavy elements. However the perspectives
of high current beams in planned a new experimental facilities at RIKEN and DUBNA
(SHE - FACTORY) give hope for success.
 An experiment with a $^{51}V$ beam is already under way at Riken

\begin{figure}[h!]
\includegraphics[width=5cm]{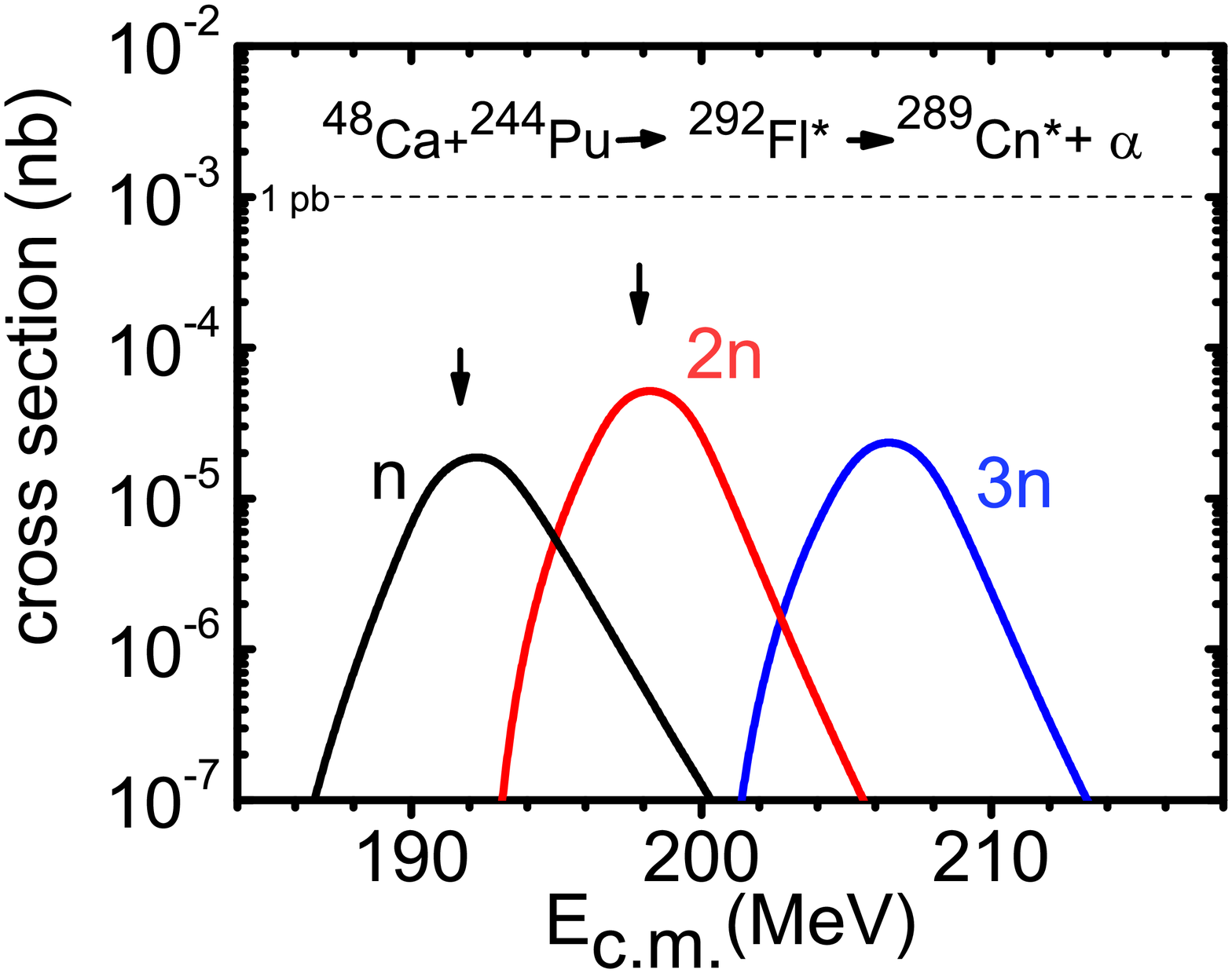}
\includegraphics[width=5cm]{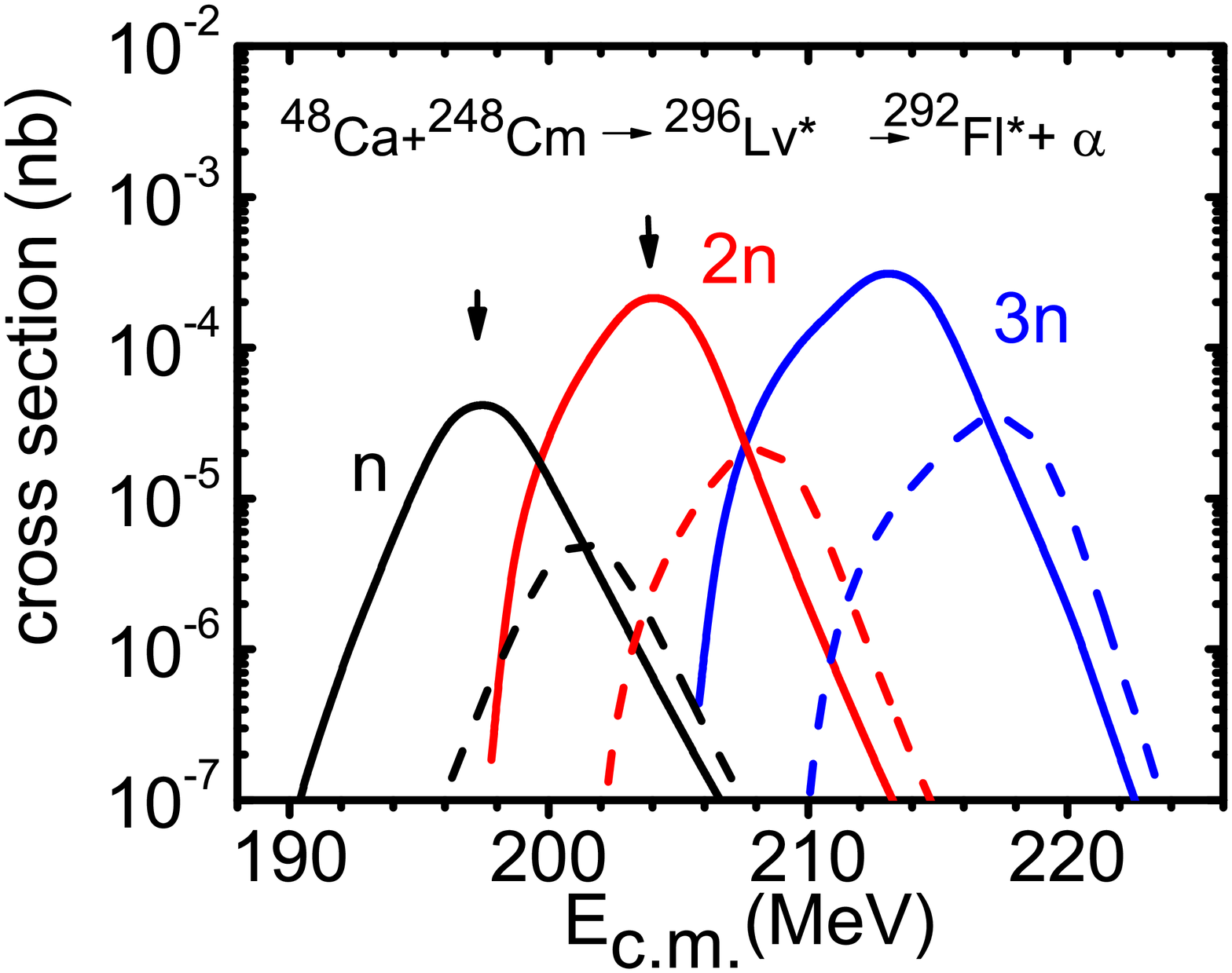}
\includegraphics[width=5cm]{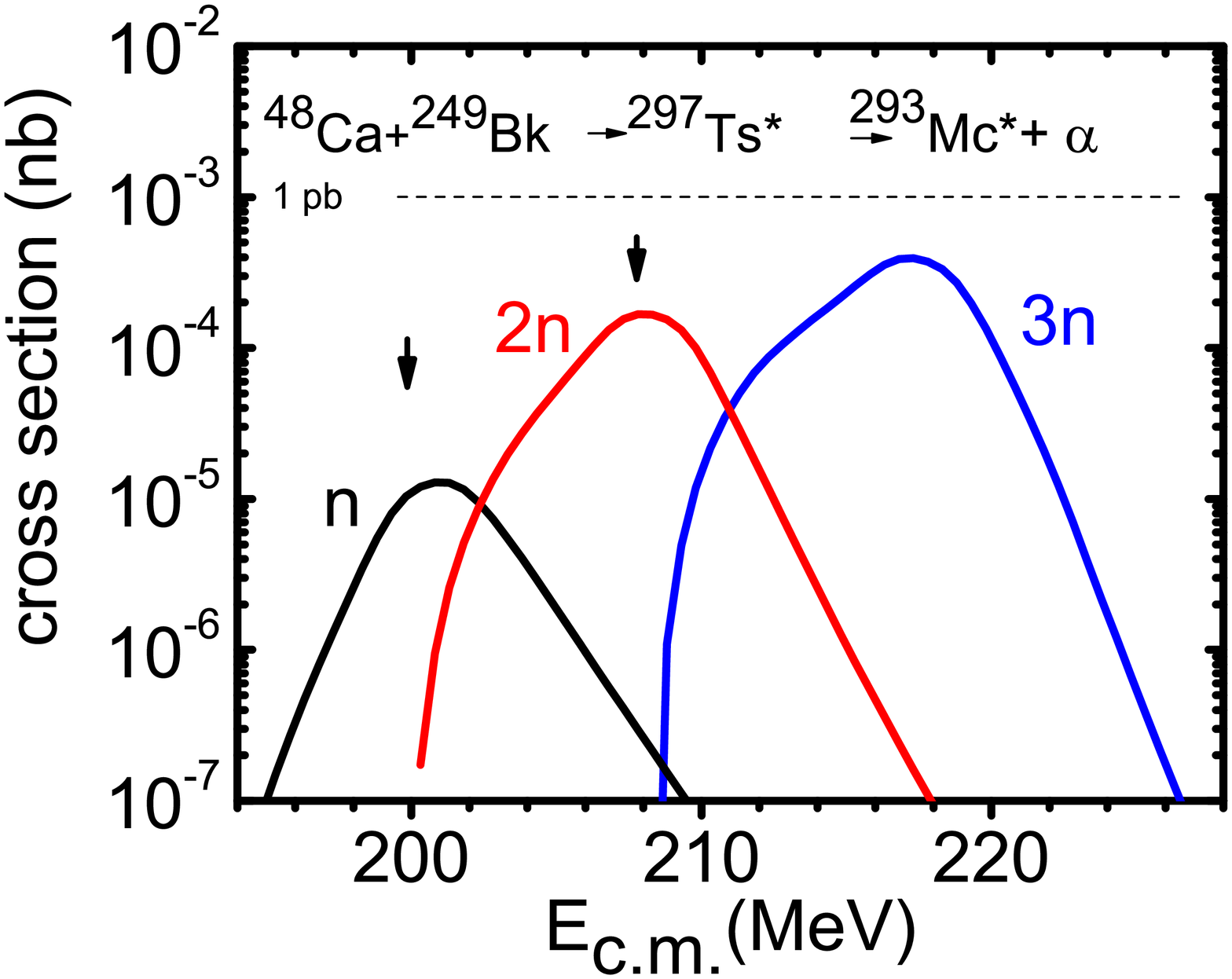}
\includegraphics[width=5cm]{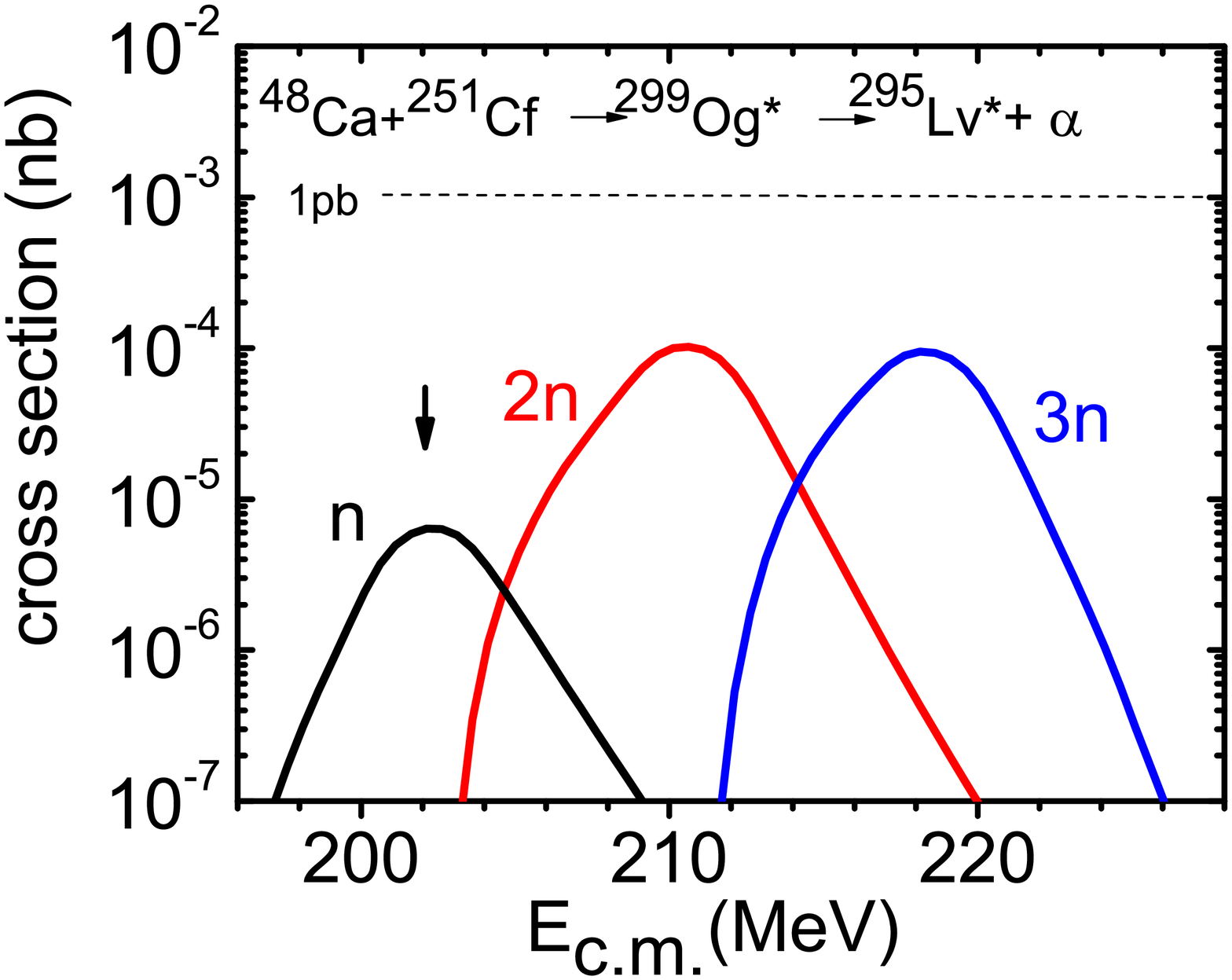}
\caption{(Color online) Cross sections for  the synthesis of superheavy nuclei in $\alpha$$xn$
fusion evaporation processes, predicted by the fusion-by-diffusion (FBD) model
with the fission barriers and ground-state masses of Kowal et al. \cite{MK,MK1} and
the systematics of the injection-point distance (see text).} \label{Fig.3}
\end{figure}
\begin{figure}[h!]
\includegraphics[width=5cm]{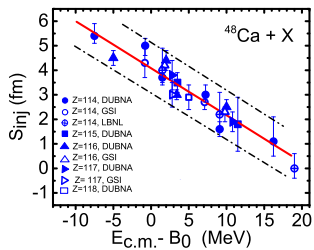}
\caption{ (Color online) Systematics of the $s_{inj}$ parameter as a function of the kinetic
energy excess $E_{c.m.}-B_{0}$ above the Coulomb barrier $B_{0}$. Solid line -
straight line approximation to experimental data, see Ref. \cite {KSW12}).
Dashed dot lines - error corridor.} \label{Fig.4}
\end{figure}

\begin{figure}[h!]
\includegraphics[width=5cm]{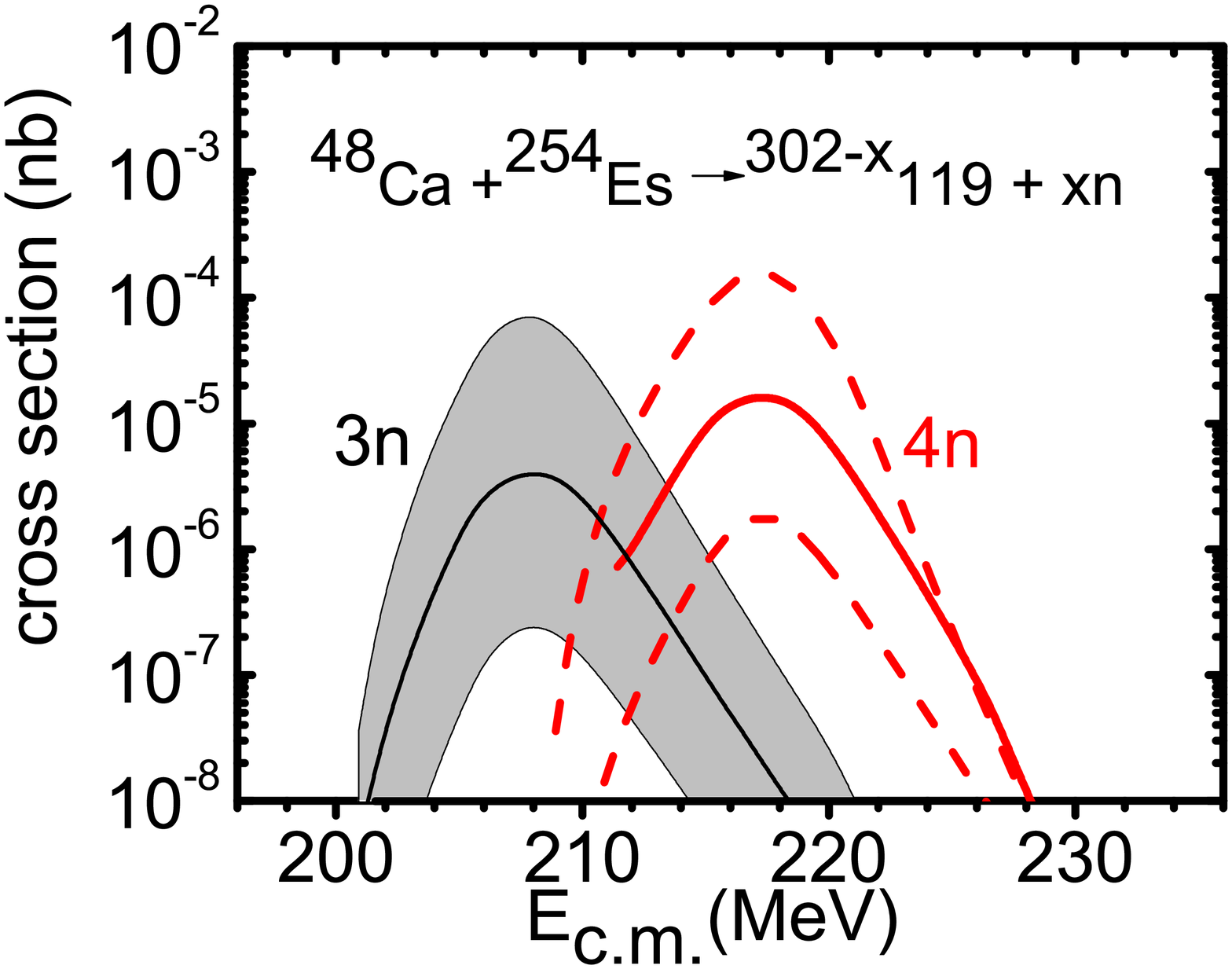}
\includegraphics[width=5cm]{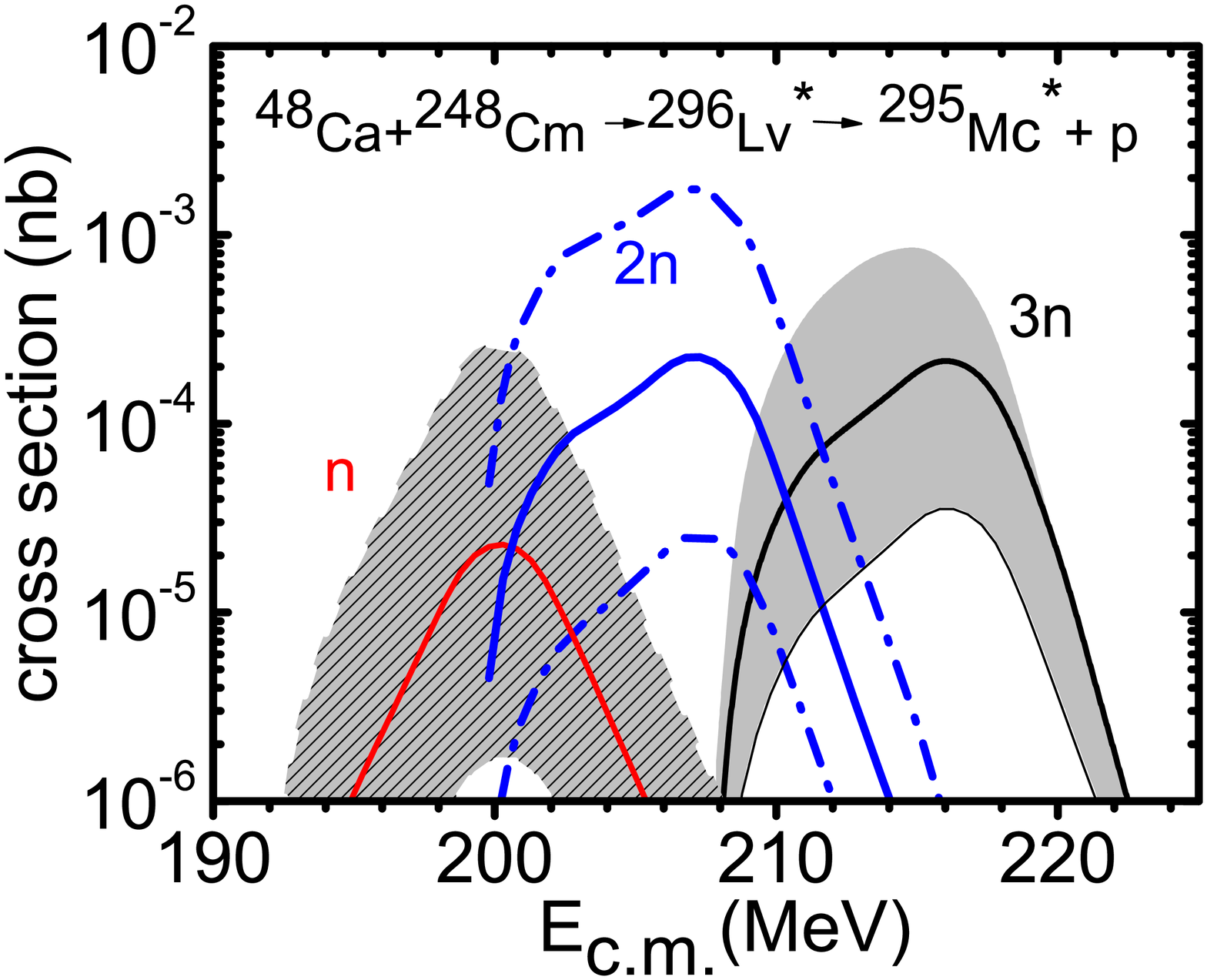}
\caption{(Color online) Excitation functions for the synthesis of superheavy nuclei in the
$^{254}Es(^{48}Ca,xn)^{302-x}119$ and $^{248}Cm(^{48}Ca,pxn)^{295-x}Mc$ fusion
evaporation processes. Solid lines correspond to calculations performed with
the straight line approximation of the $s_{inj}$. Uncertainties are defined by
the dashed-dot lines or shaded areas. } \label{Fig.5}
\end{figure}

\subsection{New isotopes of known heaviest elements}

With the perspectives of higher beam current one might expect that it will be
feasible to synthesize heavier isotopes of already known superheavy elements.
Most of these known elements were produced in the 3n or 4n fusion-evaporation
channels. Although, the 2n evaporation channels have smaller cross sections they
could lead to the synthesis of several new nuclei $^{290}Fl$, $^{294}Lv$,
$^{295}Ts$, $^{295}Og$ (see Ref. \cite{KSW12}).

In addition to the ($xn$) fusion-evaporation processes one could also consider the
fusion process in which a proton or alpha particle is evaporated (in the first step of the compound nucleus deexcitation cascade). The excited nucleus of mass
number $A_{CN-1}$ and atomic number $Z_{CN-1}$ or $A_{CN-4}$, $Z_{CN-2}$
respectively could then decay by the $xn$ cascade. Schematically:
$ P_{Zp,Ap}+T_{Zt,At}\to CN^{*}_{Z_{CN},A_{CN}}\to ER_{Z_{CN-1},A_{CN-1-x}}+p+xn $

$ P_{Zp,Ap}+T_{Zt,At}\to CN^*_{Z_{CN},A_{CN}}\to ER_{Z_{CN-2},A_{CN-4-x}}+\alpha+xn$
where, $P$ - projectile, $T$
- target, $CN^*$ - excited compound nucleus, $ER$ - evaporation residue.

To be able to predict cross sections for the above-mentioned processes, in
addition to the entry data used in calculation of $P_{surv}$ in the $xn$
processes, one needs to know the value of the Coulomb barrier between the evaporated
charged particle and the heavy nucleus with atomic number $Z = Z_{CN-1}$ or $Z =
Z_{CN-2}$. In our calculations we have used the Coulomb barrier parametrization
for protons and alpha particles proposed by Parker et. al \cite{P}
\begin {equation}
V_{p} = 0.106 (Z_{CN - 1} - 0.9) MeV 
\end {equation}
and
\begin {equation}
V_\alpha=\frac{2.88 Z_{CN-2}}{1.47 \sqrt[3]{A_{CN-4}}+4.642} MeV. 
\end {equation}
Calculations were performed for all $^{48}Ca$ 
induced reactions used to produce superheavy nuclei with atomic numbers Z between 113 and 118. Excitation
functions for reactions where new isotopes of known elements could be produced
in $pxn$ ($^{242}Pu(^{48}Ca,pxn)^{289-x}Nh$, $^{244}Pu(^{48}Ca,pxn)^{291-x}Nh$,
$^{248}Cm(^{48}Ca,pxn)^{295-x}Mc$, $^{249}Bk(^{48}Ca,pxn)^{296-x}Lv$, )
reactions are presented in Fig. 2. The last picture corresponds to reactions
on a mixed californium target $^{249-252}Cf(^{48}Ca,pxn)^{295-297}Ts$
(predictions for the synthesis of new isotopes of Og by the $xn$ evaporation
process - see Ref. \cite {Cap}).  During the experiment, which is planned at
Dubna with a new mixed californium target \cite {Ryk} in addition to
synthesizing new Og isotopes it may also be feasible to look for new isotopes
of tennesin. The cross section for synthesis of tennesin 295 in our
predictions is about 25 fb and for 296 about 7 fb. Results for the $\alpha xn$
($^{48}Ca+^{244}Pu \to^{288-x}Cn +\alpha+xn$, $^{48}Ca+^{249}Bk \to^{293-x}Mc +\alpha+xn$ and  $^{48}Ca+^{248}Cm \to^{292-x}Fl
+\alpha+xn$
 $^{48}Ca+^{251}Cf\to^{295-x}Lv +\alpha+xn$)
 reactions are shown in Fig. 3. To illustrate the influence of
the Coulomb barrier on the values of the cross sections, calculations were also
made, for selected reactions with the Coulomb barriers increased by 4 MeV
(shown as dashed lines in Fig. 2 and Fig. 3). This increase resulted in a shift
of the maximum of the excitation functions to higher energies and a decrease of
the cross section by at least one order of magnitude. The black arrows indicate
those reaction channels which lead to the formation of undiscovered new
isotopes. Although the value of the Coulomb barrier is not known exactly, the
maximum of the synthesis cross sections is in most cases above 10 fb.
Therefore, it should be possible to discover 10 new isotopes - in $pxn$
($^{287- 290}Nh$, $^{291-294}Mc$ and $^{295,296}Ts$), and 7 - in $\alpha$$xn$
($^{286,287}Cn$, $^{290,291}Fl$, $^{291,292}Mc$ and $^{294}Lv$)
fusion-evaporation reaction channels.

\section{UNCERTAINTIES}
Different theoretical models give predictions that may differ by one or even
two orders of magnitude for the same fusion-evaporation reaction. Therefore,
it is very important to estimate the uncertainties of the present
calculations. As pointed out in the description of equation (1), the synthesis
cross section consists of three factors: the partial capture cross section
$\sigma_{cap}(l) =\pi\lambdabar^2(2l + 1)T (l)$, the fusion probability
$P_{fus}(l)$, and the survival probability $P_{surv}(l)$. Each factor is
calculated within some uncertainties. In our approach, the capture cross
section should not change significantly from one system to another. The
resulting uncertainties should not be large unless deeply sub-barrier reactions
are studied.  The fusion probability depends on the asymmetry of the colliding
system and the entrance channel energy. Predictions may result in large
uncertainties of even several orders of magnitude for the unexplored region of
heavy systems. The survival probability is very sensitive to the value of the
fission barrier (a 1 MeV difference in the fission barrier height may result in
a one order of magnitude difference in the value of the cross section at each
step of the deexcitation cascade). Therefore, it is very important to do
systematic calculations using the same entry data and compare to already
measured excitation functions. In our approach there is one free
parameter - $s_{inj}$. The systematics of $s_{inj}$ as a function of the kinetic energy
excess $E_{c.m.} -B_{0}$ above the Coulomb barrier $B_{0}$, was studied using
all available experimental data for $^{48}Ca$ induced reactions. As shown in
Fig. 4 this parameter can be approximated by a straight line \cite
{KSW12}. Deviations from this line incorporate all uncertainties. The error corridor
shown by the dashed lines (see Fig. 4) should allow the
accuracy of our predictions to be estimated. As an example, two $^{48}Ca$ induced reactions are
presented in Fig. 5. Solid lines correspond to calculations performed with the
straight line approximation of the $s_{inj}$. Uncertainties are defined by the
dashed dot lines or shaded areas. Calculations were made for all studied
systems. The conclusion, based on this study, is that in our approach the
uncertainties of the predicted cross sections for $^{48}Ca$ induced reactions on
actinide targets are no better than one order of magnitude. Calculations of the
$pxn$ and $\alpha xn$ processes in $^{48}Ca$ induced reactions on actinide
targets were also performed by Hong et al. Ref. \cite {Hong}. Predictions in
most cases agree within one order of magnitude, although the model and entry
data used in the calculations are different.

\section{CONCLUSIONS}
The Fusion by Diffusion model with fission barriers and ground state masses
calculated within the Warsaw macroscopic-microscopic model was applied to
predict synthesis cross sections of superheavy nuclei in fusion-evaporation
$xn$, $pxn$ and $\alpha xn$ processes.
Anticipating the use of high current accelerators and more effective
experimental setups, calculations of the excitation functions for the synthesis
of new superheavy nuclei in the atomic number range  Z = 112 - 120  were
presented. Calculations predict the possibility of observing 21 new heaviest
nuclei with cross sections above 10 fb, among them two new elements
$^{295,296}119$ and  $^{296,297}120$. The accuracy of the predicted cross sections
was discussed.

\section*{ACKNOWLEDGEMENTS}
M.K. was co-financed by the National Science Centre under Contract No. UMO-2013/08/M/ST2/00257  (LEA COPIGAL).



\end{document}